\documentclass[a4paper,11pt]{article}
\pdfoutput=1

\usepackage{jheppub}
\usepackage{epstopdf}

\usepackage[T1]{fontenc}

\usepackage{threeparttable}
\usepackage{multirow}
\usepackage{subfigure}
\usepackage{float}
\let\oldequation\equation
\let\oldendequation\endequation
\renewenvironment{equation}
  {\linenomathNonumbers\oldequation}
  {\oldendequation\endlinenomath}

\def \jpsi {J/\psi}

\def \ee   {e^+e^-}
\def \$K$ {K}

\def \ks  {K_{S}^{0}}

\def \gev  {~\mbox{GeV}}
\def \gevc {~\mbox{GeV/$c$}}
\def \gevcc{~\mbox{GeV/$c^2$}}
\def \mev  {~\mbox{MeV}}

\begin{document}

\title{\boldmath Search for the  FCNC charmonium decay $J/\psi \to D^0 \mu^+ \mu^- + \text{c.c.}$}

\collaborationImg{\includegraphics[height=30mm,angle=90]{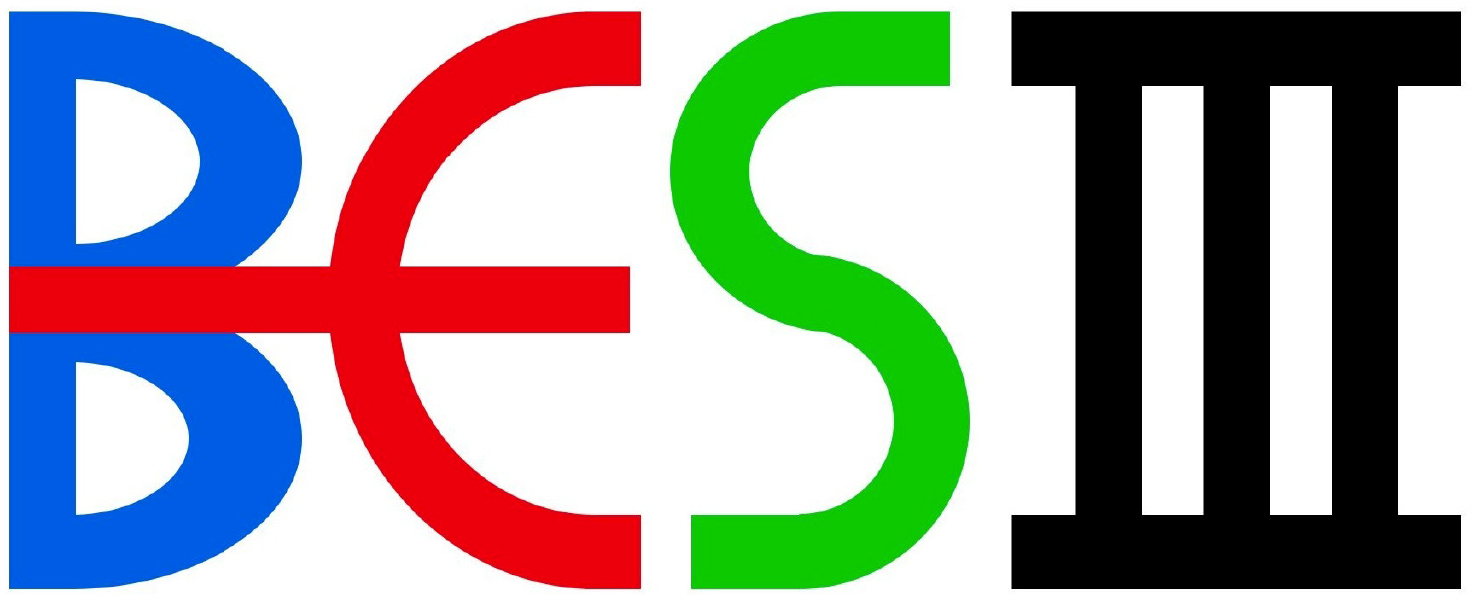}}
\collaboration{The BESIII collaboration}
\emailAdd{besiii-publications@ihep.ac.cn}

\abstract{
Based on a data sample of $(10087 \pm 44) \times 10^6$ $J/\psi$ events taken with the BESIII detector, we search for the flavor-changing neutral current charmonium decay $J/\psi \to D^{0} \mu^{+} \mu^{-} + \text{c.c.}$.
No significant signal above the background is observed, and the upper
limit on its branching fraction is set to be $\mathcal{B}(J/\psi \to
D^{0}\mu^{+}\mu^{-} + \text{c.c.} ) < 1.1 \times 10^{-7}$ at the 90\%
confidence level. This marks the first search for a flavor-changing
neutral current charmonium decay involving muons in the final state.
}


\arxivnumber{}

\maketitle
\flushbottom


\section{INTRODUCTION}
\label{sec:introduction}
\hspace{1.5em} In the Standard Model~(SM) of particle physics, the
process of flavor-changing neutral currents~(FCNC)~\cite{FCNC_new1,FCNC_new2} is
prohibited at the tree level due to Glashow-Iliopoulos-Maiani
mechanism~\cite{GIM}. So FCNC transitions such as $c \to u$ manifest
solely through loop-level interactions, which result in low branching
fractions~(BFs) of the corresponding decays.  Nevertheless, various
new physics~(NP) models beyond the SM, such as the Top-Color
model~\cite{Topcolor}, the supersymmetric extensions of the SM with or without
R-parity violation~\cite{supersymmetric}, and the two-Higgs doublet
model~\cite{twoHiggs}, suggest these BFs may be enhanced to exceed the
SM prediction.  Some dark sector particles, like QCD
axion~\cite{dakersector1} or massless dark photon~\cite{dakersector2},
with flavor-violating couplings, can also contribute
to the FCNC process.  Any observation of such rare FCNC processes
beyond the SM prediction will be a clear signal for NP~\cite{goodprobe2}.  Experimentally, FCNC processes have
been extensively studied in the $B$ meson sector via $b \to s$
transitions~\cite{B1, B2}, and in the $D_{(s)}$ meson sector through
$c \to u$
transitions~\cite{BESIII:2018hqu,LHCb:2020car,BESIII:2021slf,BESIII:2024nrw,dakersector3,BESIII:2022vrr,dakersector4}.
In contrast, measurements of the  FCNC charmonium decays are relatively
limited.

The charmonium such as $\jpsi$, primarily decaying via strong and electromagnetic interactions, has been extensively studied for decades.
Given its mass below the $D^{0}\bar{D^{0}}$ mass threshold, $\jpsi$ could not decay into $D^{0}\bar{D^{0}}$ pairs.
Instead, it can decay weakly into a single $D$ meson with a minuscule
BF, which has been searched for by BES and BESIII collaborations~\cite{Ablikim:2019hff, Li:2024moj, bes:2008, bes3:2014xbo, bes3:2014,  bes3:2021, jpsi2Dmunu, BESIII:2022ibp, BESIII:2023qpx,jpsi2D0ee,BESIII:gammaD0}.
The searching for the FCNC charmonium decays is crucial to study the charmonium
weak decay mechanism, and provides an opportunity to study the non-perturbative QCD effects.
Compared to the studies of the FCNC $D$ decays, the studies of the
$\jpsi$ FCNC decays benefit from a larger data sample at BESIII~\cite{bes3:totJpsiNumber}.
The FCNC decay $\jpsi \to D^0 l^+ l^-$ is expected to have a BF on the order of $10^{-13}$ in the SM~\cite{FCNC2}.
So far, several FCNC processes have been probed in the charmonium, such as $\jpsi \to D^0 e^+ e^-$~\cite{jpsi2D0ee}, $\psi(3686)\to D^0e^+e^-$~\cite{jpsi2D0ee} and $\jpsi \to \gamma D^0$~\cite{BESIII:gammaD0}, but no conclusive signals have been observed, as shown in Table~\ref{tab:jpsi decay}.
The $\jpsi$ FCNC decays involving muons in the final state have not yet been measured.

\begin{table*}[!htbp]
\caption{Experimental results of the charmonium FCNC decays. The decay modes, the total number of $\jpsi$ or $\psi(3686)$ events, and the upper limit~(UL) on the BF at the 90\% confidence level~(C.L.) are listed.}
\setlength{\abovecaptionskip}{1.2cm}
\setlength{\belowcaptionskip}{0.2cm}
\begin{center}
\footnotesize
\vspace{-0.0cm}
\begin{tabular}{l|llll}
\hline \hline
		Experiment & Decay mode & $N_{\jpsi}$ or $N_{\psi(3686)}$ & UL & Year\\
		\hline
        BESIII & $J/\psi\to D^0e^+e^-$ & $1310.6\times10^{6}$ & $8.5\times10^{-8}$ & 2017 \cite{jpsi2D0ee}\\
        BESIII & $\psi(3686)\to D^0e^+e^- $ & $447.9\times10^{6}$ & $1.4\times10^{-7}$ & 2017 \cite{jpsi2D0ee}\\
        BESIII & $J/\psi\to \gamma D^0 $ & $10087\times10^{6}$ & $9.1\times10^{-8}$ & 2024 \cite{BESIII:gammaD0}\\

\hline \hline
\end{tabular}
\label{tab:jpsi decay}
\end{center}
\end{table*}
\vspace{-0.0cm}

\begin{figure*}[htbp]
\centering
	\setlength{\abovecaptionskip}{-1pt}
	\setlength{\belowcaptionskip}{10pt}
	\includegraphics[width=0.5\textwidth]{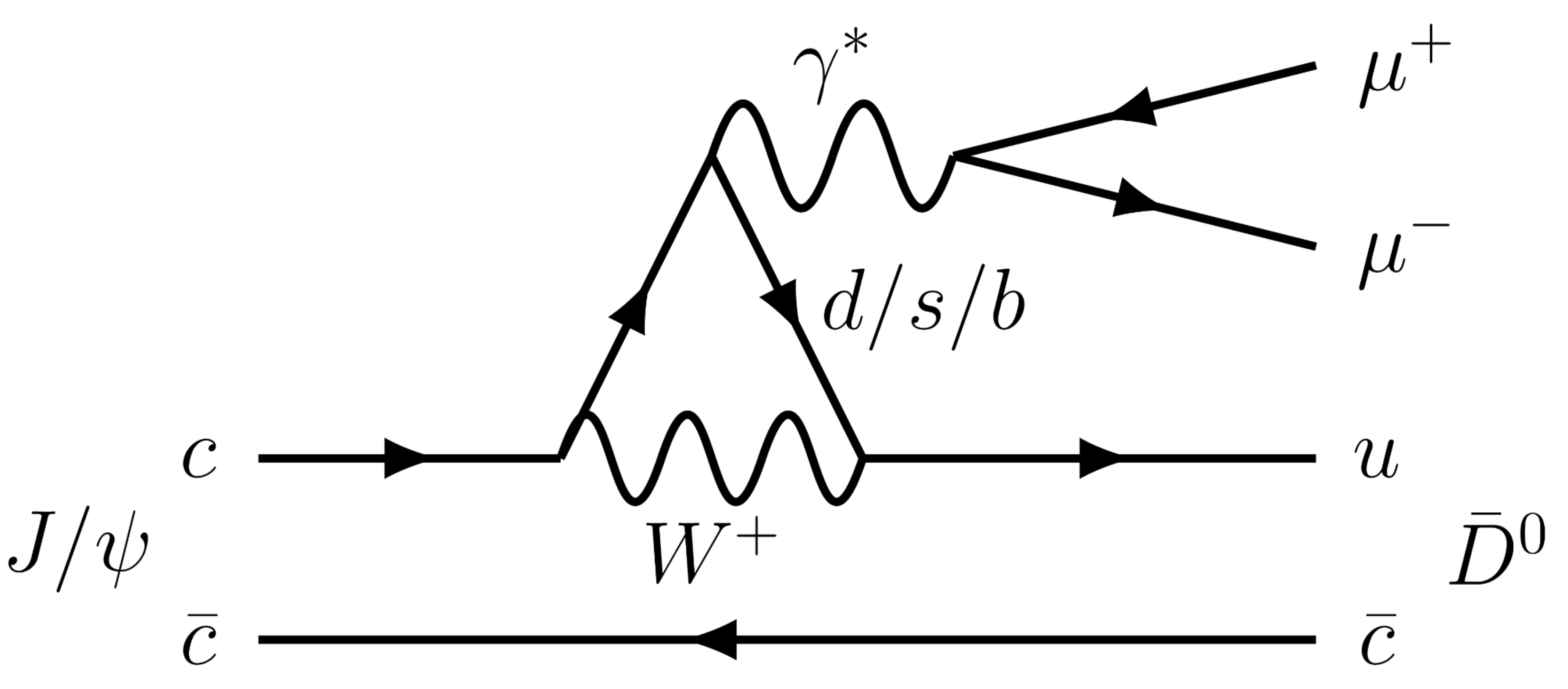}
	\caption{Feynman diagram for the $\jpsi\to \bar{D}^{0}\mu^+\mu^-$ decay in the SM.}
\label{fig:feynman}
\end{figure*}

In this paper, we report the first search for the FCNC decay
$\jpsi \to D^{0} \mu^{+}\mu^{-}$ based on $(10087\pm44)\times10^{6}$
$\jpsi$ events.  This decay is a typical FCNC process that can occur
through possible Feynman diagrams such as the one shown in
Figure~\ref{fig:feynman}. 
Throughout this paper, charge-conjugate processes are implied.

\section{BESIII DETECTOR AND MONTE CARLO SIMULATION}
\label{sec:detector}
\hspace{1.5em}
The BESIII detector~\cite{Ablikim:2009aa} records symmetric $e^+e^-$ collisions
provided by the BEPCII storage ring~\cite{Yu:IPAC2016-TUYA01}
in the center-of-mass energy range from 1.84 to 4.95~GeV,
with a peak luminosity of $1.1 \times 10^{33}\;\text{cm}^{-2}\text{s}^{-1}$
achieved at $\sqrt{s} = 3.773\;\text{GeV}$.
BESIII has collected large data samples in this energy region~\cite{Ablikim:2019hff}. The cylindrical core of the BESIII detector covers 93\% of the full solid angle and consists of a helium-based
 multilayer drift chamber~(MDC), a time-of-flight
system~(TOF), and a CsI(Tl) electromagnetic calorimeter~(EMC),
which are all enclosed in a superconducting solenoidal magnet
providing a 1.0~T magnetic field.
The magnetic field was 0.9~T in 2012, which affects 11\% of the total $J/\psi$ data.
The solenoid is supported by an
octagonal flux-return yoke with resistive plate counter muon
identification modules interleaved with steel.
The charged-particle momentum resolution at $1~{\rm GeV}/c$ is
$0.5\%$, and the
${\rm d}E/{\rm d}x$
resolution is $6\%$ for electrons
from Bhabha scattering. The EMC measures photon energies with a
resolution of $2.5\%$ ($5\%$) at $1$~GeV in the barrel (end cap)
region. The time resolution in the plastic scintillator TOF barrel region is 68~ps, while
that in the end cap region was 110~ps. The end cap TOF
system was upgraded in 2015 using multigap resistive plate chamber
technology, providing a time resolution of
60~ps,
which benefits 87\% of the data used in this analysis~\cite{etof, etof2, etof3}.

The inclusive Monte Carlo~(MC) sample and signal MC sample simulated
data samples produced with a {\sc geant4}-based~\cite{geant4} software
package, which includes the geometric description of the BESIII
detector~\cite{bes:unity, geo1, geo2} and the detector response, are
used to determine detection efficiencies and to estimate the
background contributions. The simulation models the beam energy spread and initial state radiation in the $e^+e^-$ annihilations with the generator {\sc kkmc}~\cite{ref:kkmc, ref:kkmc2}.
All particle decays are modeled with {\sc evtgen}~\cite{ref:evtgen, ref:evtgen2} using BFs
either taken from the Particle Data Group~(PDG)~\cite{pdg:2024}, when available, or otherwise estimated with {\sc lundcharm}~\cite{ref:lundcharm, ref:lundcharm2}. Final state radiation~(FSR) from charged final state particles is incorporated using the {\sc photos} package~\cite{photos2}.

\section{EVENT SELECTION AND DATA ANALYSIS}
\label{sec:analysis}
\hspace{1.5em}
The data and simulated MC samples used in this analysis are reconstructed with the BESIII offline software system~\cite{besofflinesoftware1, besofflinesoftware}.
We explore the decay $\jpsi \to D^{0}\mu^{+}\mu^{-}$, utilizing three tag modes to reconstruct $D^0$ meson: $D^{0} \to K^{-}\pi^{+}$~(Mode I), $D^{0} \to K^{-}\pi^{+}\pi^{0} $ with $\pi^0\to\gamma\gamma$~(Mode II), and $D^{0} \to K^{-}\pi^{-}\pi^{+}\pi^{+} $~(Mode III), which have large BFs.

The charged tracks are expected to be detected within the MDC,
confined to a polar angle ($\theta$)  range of $|\cos\theta|<0.93$, where $\theta$ is defined with respect to the $z$-axis,
which is the symmetry axis of the MDC.
The closest approach of tracks to the interaction point~(IP) is required to be less than 10~cm along the $z$-axis and less than 1~cm in the transverse plane.
For Modes~I, II, and III, the numbers of good charged tracks in each event are required to be 4, 4, and 6, respectively, with a net charge equal to 0.

Particle identification~(PID) is applied by combining the measurements
of $dE/dx$ in the MDC and the flight time in the TOF to calculate the
likelihoods $\mathcal{L}(h)~(h=p,K,\pi)$ for each hadron $h$ hypothesis.
For the $\mu$ hypothesis, information from the EMC is included.
The $\pi$ candidates must satisfy $\mathcal{L}(\pi)>0$ and $\mathcal{L}(\pi)>\mathcal{L}(K)$.
The $K$ candidates need to have $\mathcal{L}(K)>0 $ and $ \mathcal{L}(K)>\mathcal{L}(\pi)$.
The $\mu$ candidates are required to satisfy $\mathcal{L}(\mu)>0.001 $, $ \mathcal{L}(\mu)>\mathcal{L}(e) $, and $ \mathcal{L}(\mu)>\mathcal{L}(K)$.
The mass of the $\mu$ lepton is close to that of the $\pi$ meson, causing $\mu / \pi$ separation more difficult.
This challenge is further intensified by the low momentum of the $\mu$
in the signal events, for which the MUC offers very limited assistance
for $\mu / \pi$ identification due to low efficiency.
The $\mu$ candidate selection is therefore based on the $\mu$ energy deposit in
the EMC~($E_{\mu}$) by requiring  $0.11\gev < E_{\mu} < 0.25\gev$.

Photon candidates are identified using isolated showers in the EMC.
The deposited energy of each shower $E_{\gamma}$ must satisfy
$E_{\gamma} > 25~\mev$ in the barrel region
($|\text{cos}\theta|<0.8$), and $E_{\gamma} > 50~\mev$ in the end cap
region ($0.86 < |\cos\theta| < 0.92$). To exclude showers that
originate from charged tracks, the angle subtended by the EMC shower
and the position of the closest charged track at the EMC must be
greater than 10 degrees as measured from the IP. To suppress
electronic noise and showers unrelated to the event, the difference
between the EMC time and the event start time is required to be within
[0, 700]\,ns.  Events with at least two photon candidates are retained
in the selection of the $D^{0} \to K^{-}\pi^{+}\pi^{0}$ tag mode. In
this case, all good photon candidates are looped over, and the
$\pi^{0}$ candidate is selected as the combination with the minimum
$\chi^{2}(\gamma\gamma)$ in a kinematic fit~\cite{Yan:2010zze} of the
two photons to the $\pi^{0}$ nominal mass~\cite{pdg:2024}.



\begin{figure*}[htbp] \centering
	\setlength{\abovecaptionskip}{-1pt}
	\setlength{\belowcaptionskip}{10pt}
        \subfigure[]
        {\includegraphics[width=0.32\textwidth]{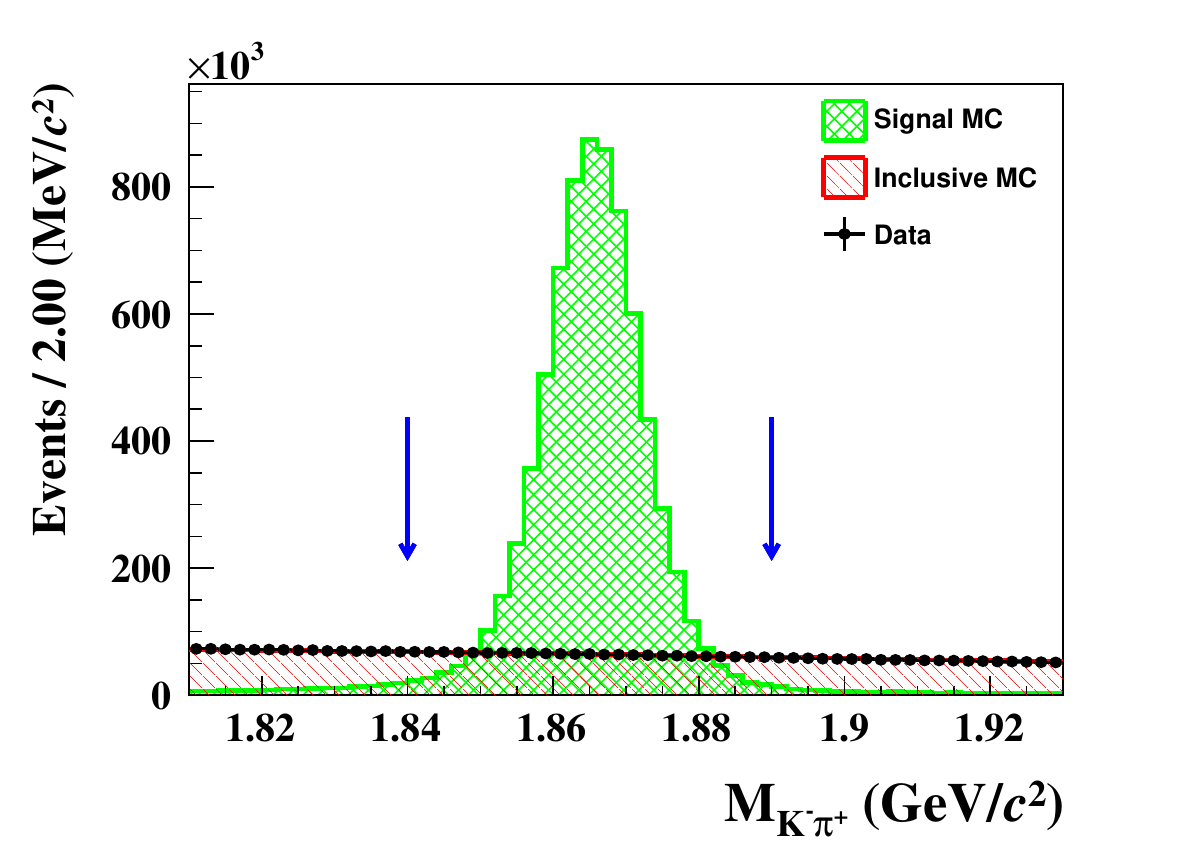}}
        \subfigure[]
        {\includegraphics[width=0.32\textwidth]{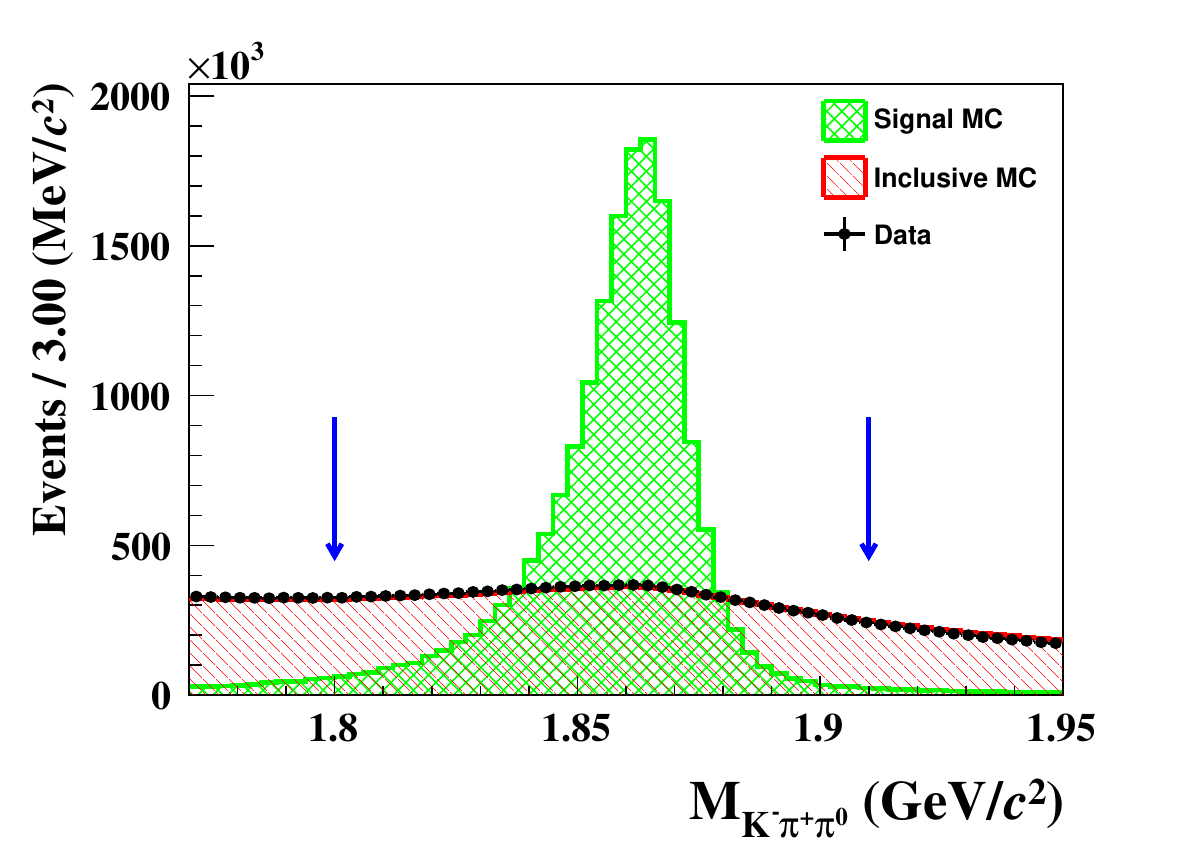}}
        \subfigure[]
        {\includegraphics[width=0.32\textwidth]{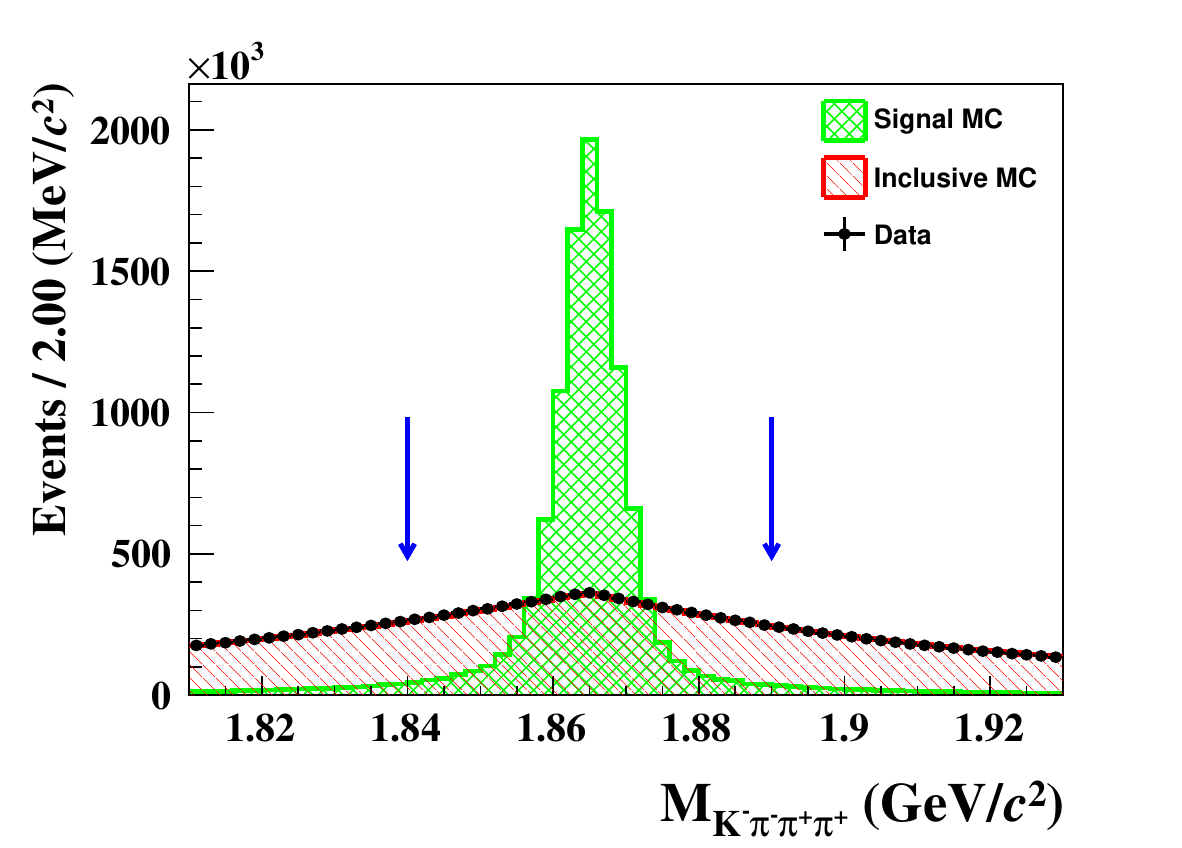}}

        \subfigure[]
        {\includegraphics[width=0.32\textwidth]{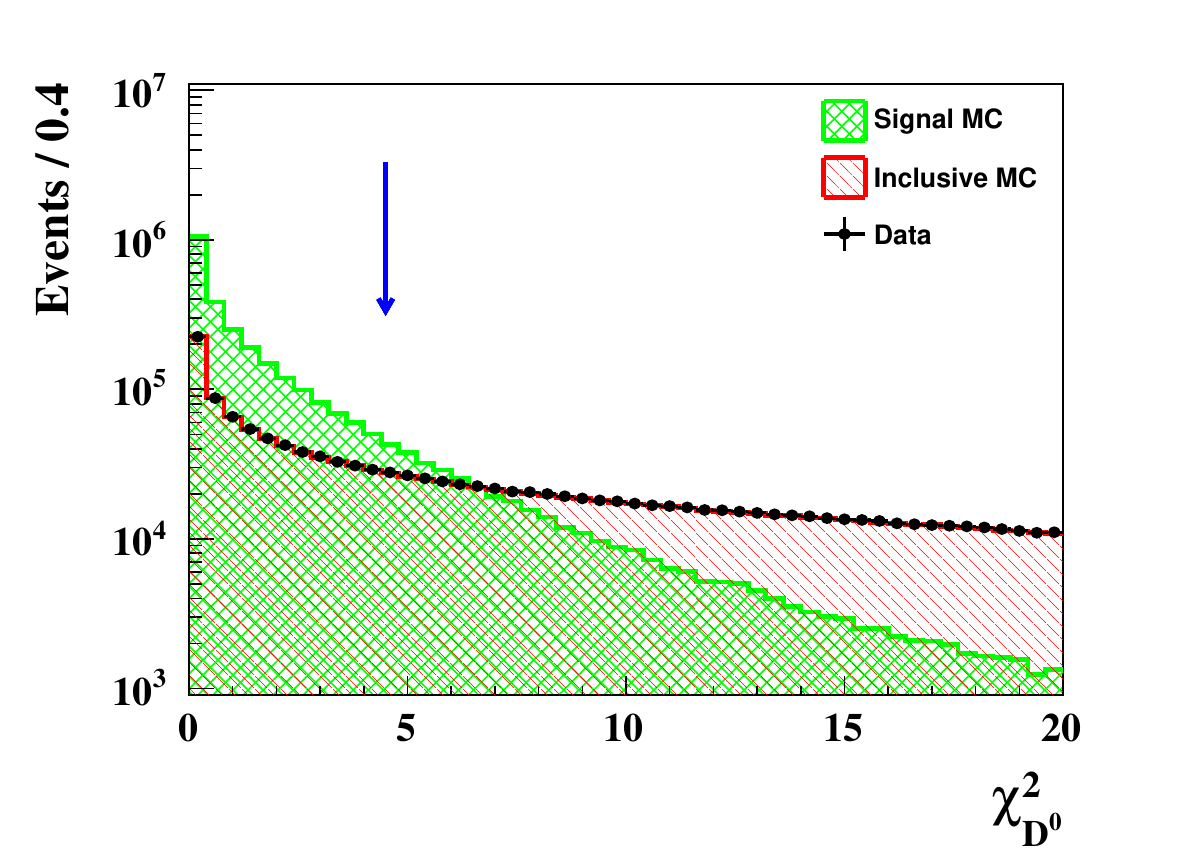}}
        \subfigure[]
        {\includegraphics[width=0.32\textwidth]{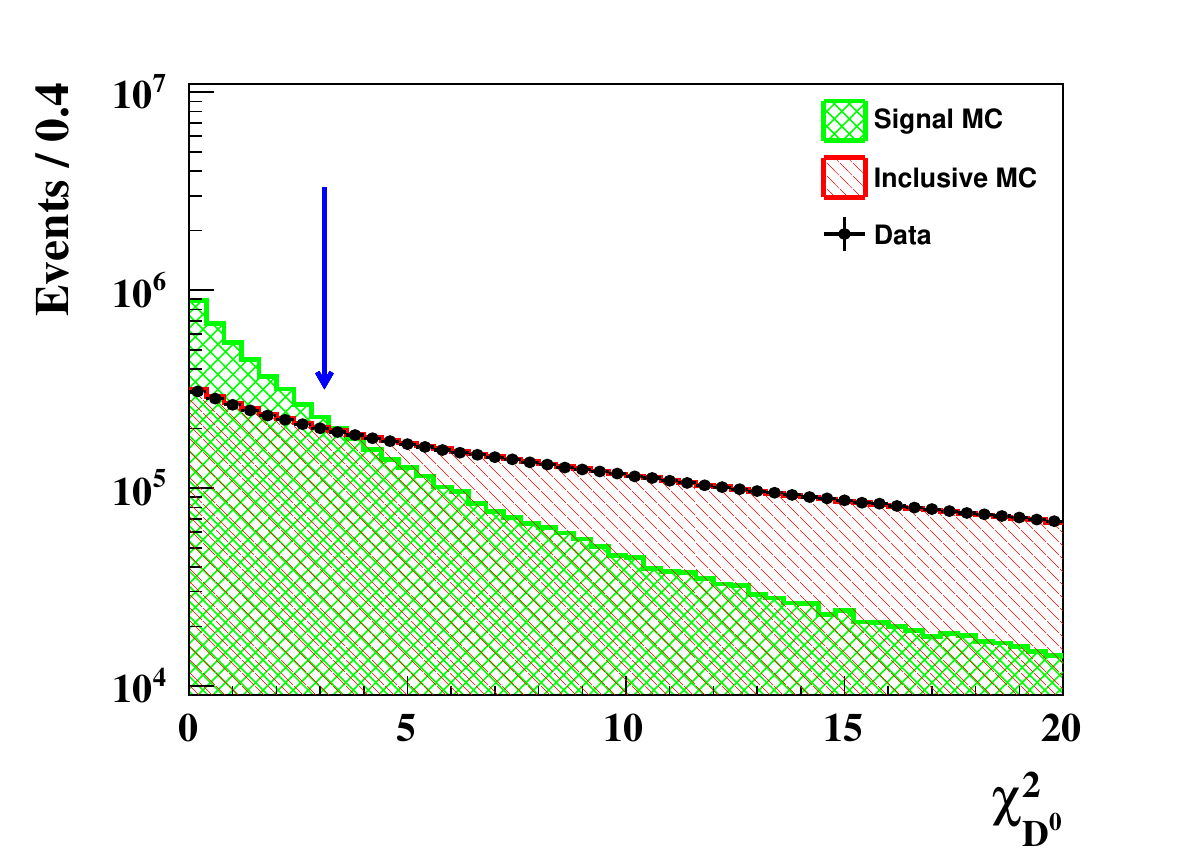}}
        \subfigure[]
        {\includegraphics[width=0.32\textwidth]{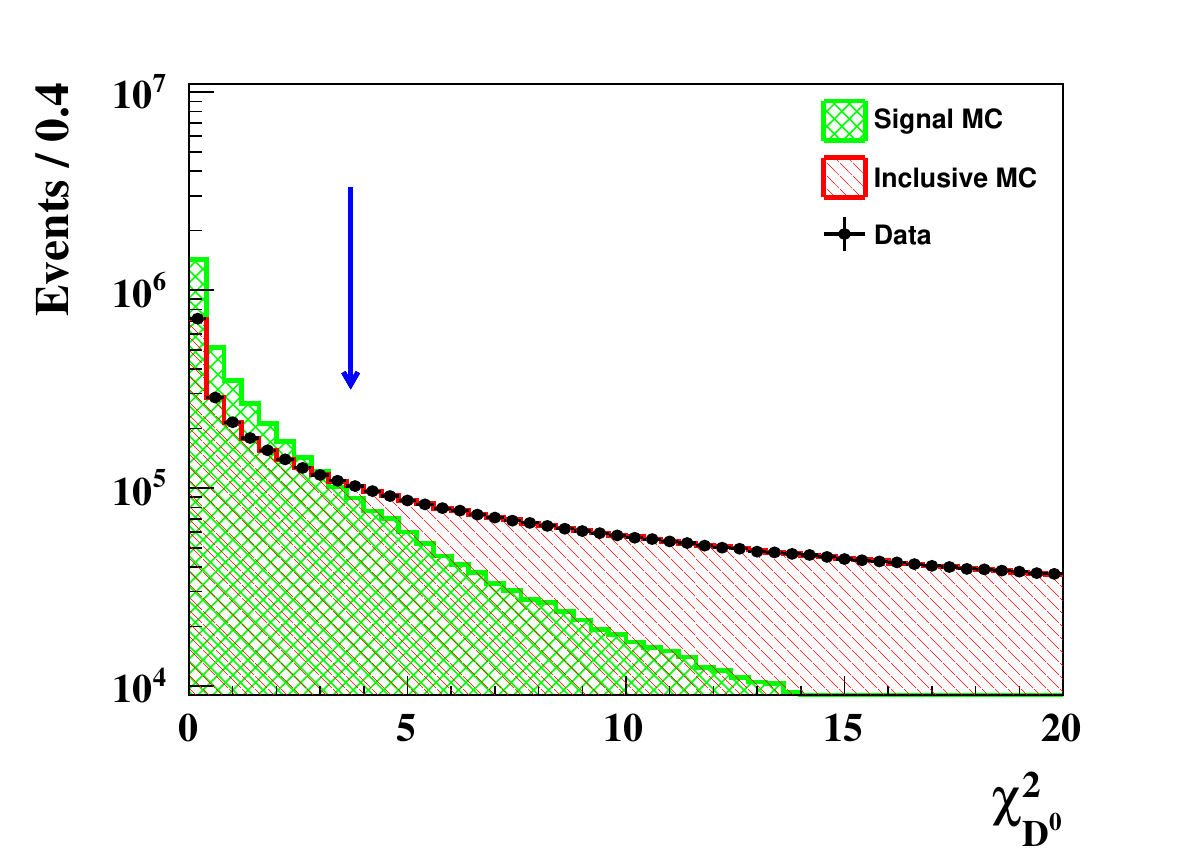}}

	\caption{
 The invariant mass $M_{D^{0}}$ distributions for $D^{0} \to K^{-}\pi^{+}$~(a), $D^{0} \to K^{-}\pi^{+}\pi^{0}$~(b), $D^{0} \to K^{-}\pi^{+}\pi^{+}\pi^{-} $~(c), along with the corresponding $\chi_{D^0}^2$ distributions for $D^{0} \to K^{-}\pi^{+}$~(d), $D^{0} \to K^{-}\pi^{+}\pi^{0}$~(e), $D^{0} \to K^{-}\pi^{+}\pi^{+}\pi^{-} $~(f).
 The green shaded histogram shows the signal MC sample scaled to $\mathcal{B}(\jpsi\to D^{0}\mu^{+}\mu^{-}) = 4.0\times10^{-2}$ for the $M_{D^{0}}$ distributions and $\mathcal{B}(\jpsi\to D^{0}\mu^{+}\mu^{-}) = 1.5\times10^{-2}$ for the $\chi_{D^0}^2$ distributions.
 The red histogram is the inclusive MC sample,
 and the black dots with error bars are data,
 and the blue arrows indicate the cut region or position.
 Due to the large amount of background, the error bars on the data are small.
 }
	\label{fig:D0}
\end{figure*}

All combinations of final state candidates are subject to a kinematic
fit~\cite{Yan:2010zze} constraining $M_{K^-\pi^+}$,
$M_{K^-\pi^+\pi^0}$, $M_{K^-\pi^+\pi^+\pi^-}$ to the nominal $D^0$
mass. The $D^{0}$ candidate is selected as the combination with the
minimum $\chi_{D^0}^2$ for each tag mode.  
Then the invariant mass of $D^0$ before the kinematic fit is required to be in the region $[1.84,1.89]\gevcc$, $[1.80,1.91]\gevcc$, and $[1.84,1.89]\gevcc$ for Modes~I, II, and III, respectively, as shown in Figure~\ref{fig:D0}~(a),~(b), and~(c).
To further suppress the non-$D^0$ background, the $\chi_{D^0}^2$ after the kinematic fit is required to be less than $4.5$, $3.1$, and $3.7$ for Modes~I, II, and III, respectively, as shown in Figure~\ref{fig:D0}~(d),~(e), and~(f).
The $\chi_{D^0}^2$ requirements have been optimized by maximizing the figure of merit~(FOM), defined as $\rm{FOM} = \frac{S}{a/2 + \sqrt{B}}$ with $a = 3$~\cite{Punzi:2003bu}, 
where $S$~($B$) represents the number of events in the signal~(inclusive) MC sample under the scanned selection of $\chi_{D^0}^2$.

\vspace{-0.0cm}
\vspace{-0.0cm}

After selecting the $D^0$ candidate, the remaining two charged particles must satisfy the muon PID criteria described above.
To reduce the background with more missing particles, the missing
momentum $\vec{p}_{\rm{miss}}$ in the center-of-mass frame of $\jpsi$ is employed, which is defined as
\begin{eqnarray}
\vec{p}_{\rm{miss}} = \vec{p}_{D^{0}}+\vec{p}_{\mu^{+}}+\vec{p}_{\mu^{-}},
\end{eqnarray}
 where $\vec{p}_{D^{0}}$, $\vec{p}_{\mu^{+}}$, and $\vec{p}_{\mu^{-}}$ are the momenta of $D^{0}$, $\mu^{+}$ and $\mu^{-}$ in the center-of-mass frame, respectively.
The $|\vec{p}_{\rm{miss}}|$ is required to be less than $0.05 \gevc$.

After the above selection criteria, several types of hadronic backgrounds survive in the inclusive MC sample.
Most background events pass the selections due to the $\pi/\mu$, $K/\mu$, or $K/\pi$ mis-identification.
To reduce the background from $K/\mu$ or $K/\pi$ mis-identification, the mass of $K$ calculated by recoiling other particles in the final state, denoted as $M^{\rm{Recoil}}_{\pi\mu\mu}$, $M^{\rm{Recoil}}_{\pi\pi^0\mu\mu}$, $M^{\rm{Recoil}}_{\pi\pi\pi\mu\mu}$ for Modes~I, II, and III, respectively, is required be in the region of $[0.43,0.56]\gevcc$.
Further suppression of the remaining backgrounds for different tag  modes are described below.

For Mode I, the remaining dominant backgrounds include $J/\psi \to K^- \ks \pi^+ $ with $\ks \to \pi^+ \pi^-$ and $J/\psi \to \pi^+ \pi^+ \pi^- \pi^- $. To reduce the background from $J/\psi \to K^- \ks \pi^+ $, the invariant mass of any combination of $\pi^+ \pi^-$, $M_{\pi^+ \pi^-}$, is required to lie outside the range $[0.45,0.55]\gevcc$. To suppress the backgrounds from $J/\psi \to \pi^+ \pi^+ \pi^- \pi^- $,  we require the invariant mass $ M_{4\pi} $ to be less than $3.05 \gevcc$.
The residual backgrounds include $\jpsi \to \pi^+\pi^+\pi^-\pi^-$, $\jpsi \to \pi^+\pi^+\pi^-\pi^-\pi^0$, and so on.


For Mode II, the remaining dominant backgrounds include $J/\psi \to K^- \ks \pi^+\pi^0$ with $\ks \to \pi^+ \pi^-$, $J/\psi \to K^+ K^- \pi^+ \pi^- $ with $K^\pm \to \pi^\pm\pi^0$, and $J/\psi \to \pi^+ \pi^+ \pi^- \pi^- \pi^{0}$.
To reduce the background from $J/\psi \to K^- \ks \pi^+ \pi^0$,
the same method as above has been applied with a slightly different $\ks$ veto window of $[0.45,0.54]\gevcc$.
To suppress the backgrounds from $J/\psi \to \pi^+ \pi^+ \pi^- \pi^- \pi^0$,  we require the invariant mass $ M_{4\pi\pi^0} $ to be less than $3.03 \gevcc$.
To suppress the backgrounds from $J/\psi \to K^+ K^- \pi^+ \pi^- $ with $K^\pm \to \pi^\pm\pi^0$,
we reject candidates with $|M_{\pi^\pm \pi^0} - M_{K^\pm}| \in  [0.47,0.51] \gevcc$.
The residual backgrounds include $\jpsi \to \pi^+\pi^+\pi^-\pi^-\pi^0\pi^0$, $\jpsi \to \pi^+\pi^+\pi^-\pi^-\pi^0$, and so on.

For Mode III, the remaining dominant backgrounds include
$\jpsi \to K^+\pi^+\pi^-\pi^{-} \ks $ with $\ks \to \pi^+ \pi^-$ and
$J/\psi \to K^+ K^- \pi^+ \pi^- $ with $K^\pm \to \pi^\pm \pi^+\pi^-$.
To remove $\jpsi \to K^+\pi^+\pi^-\pi^{-} \ks$, we again apply a similar $\ks$
veto window of $[0.44,0.54]\gevcc$ as above.  To suppress the
backgrounds from $J/\psi \to K^+ K^- \pi^+ \pi^- $ with
$K^\pm \to \pi^\pm \pi^+\pi^-$, we reject candidates with
$|M_{\pi^\pm \pi^+\pi^-} - M_{K^\pm}| \in [0.47,0.51] \gevcc$.  The
residual backgrounds include
$\jpsi \to \pi^+\pi^+\pi^+\pi^-\pi^-\pi^-\pi^0$,
$\jpsi \to K^+\pi^+\pi^+\pi^-\pi^-\pi^-$, and so on.

The invariant mass $M_{D^{0}\mu^{+}\mu^{-}}$ is defined by
\begin{eqnarray}
    \label{eq:jpsimass}
    M_{D^{0}\mu^{+}\mu^{-}}=\sqrt{ (E_{D^{0}} + E_{\mu^{+}} + E_{\mu^{-}})^2-|\vec{p}_{D^{0}} + \vec{p}_{\mu^{+}} + \vec{p}_{\mu^{-}}|^2},
\end{eqnarray}
where $E_{D^{0}},~E_{\mu^{+}},~E_{\mu^{-}}$ are calculated by $E = \sqrt{M^2 + |\vec{p}|^2}$, with $M$ representing the corresponding nominal mass~\cite{pdg:2024}, $\vec{p}_{D^{0}}$ is the momentum of $D^0$ after the kinematic fit, and $\vec{p}_{\mu^{+}},~\vec{p}_{\mu^{-}}$ are the momenta measured by the MDC.
The $M_{D^{0}\mu^{+}\mu^{-}}$ distribution of signal events is
expected to peak around the $\jpsi$ nominal mass, so the signal region is defined as $3.05\gevcc <M_{D^{0}\mu^{+}\mu^{-}} < 3.15\gevcc$.
The signal efficiencies, evaluated using signal MC samples, are $(20.23 \pm 0.05)\%$, $(7.07\pm 0.03)\%$, and $(6.69\pm 0.03)\%$ for Modes~I, II, and III, respectively.

\section{EXTRACTION OF SIGNAL YIELDS}
\label{sec:result}
\begin{figure*}[htbp]
\centering
	\setlength{\abovecaptionskip}{-1pt}
	\setlength{\belowcaptionskip}{10pt}
	\includegraphics[width=0.99\textwidth]{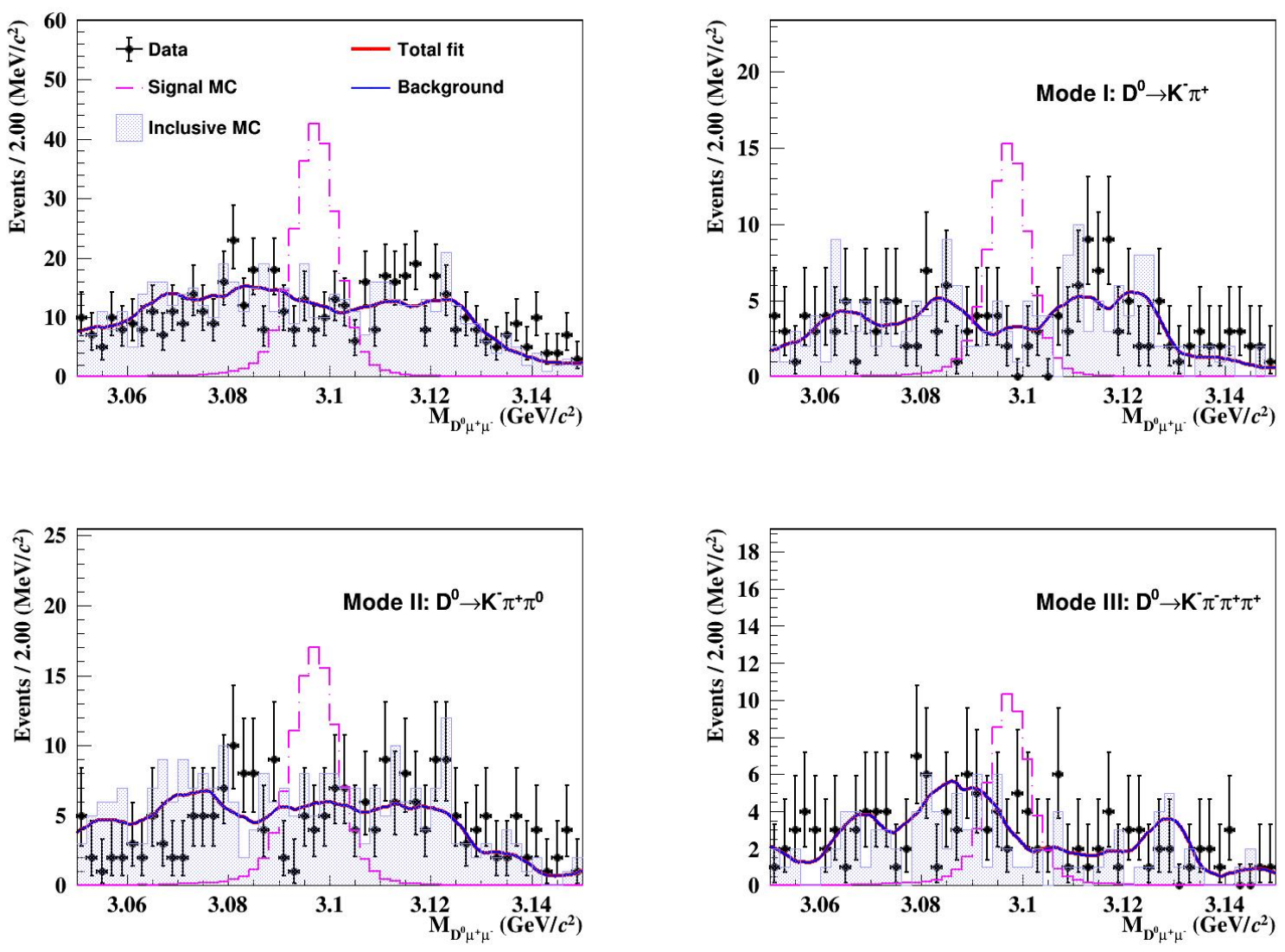}
	\caption{
The distributions of $M_{D^{0}\mu^{+}\mu^{-}}$ for $\jpsi\to D^{0}\mu^+\mu^-$ of the selected candidates in data, signal MC sample, and inclusive MC sample.
The simultaneous fit result is shown in the top left, and the individual fit result for each mode is shown in the other three sub-figures, respectively.
The black dots with error bars are data,
the magenta dotted-dashed line shows the shape of signal MC sample scaled to $\mathcal{B}(\jpsi\to D^{0}\mu^{+}\mu^{-}) = 1.0\times10^{-6}$ for the three tags.
The blue shaded histogram is the inclusive MC sample,
the red line is the fit result,
and the blue solid lines is the fitted background.
}
	\label{fig:fit}
\end{figure*}

\hspace{1.5em}
An unbinned maximum likelihood fit is performed on the
$M_{D^{0}\mu^{+}\mu^{-}}$ distribution to extract the signal yields.
The fit function is defined as
\begin{eqnarray}
    F_{\rm{fit}} &=& \sum_i (N_{\rm{sig,fit}}  \frac{\mathcal{B}_{\rm{inter},\it{i}}  \epsilon_{\rm{sig},\it{i}}}{\sum_j \mathcal{B}_{\rm{inter},\it{j}}  \epsilon_{\rm{sig},\it{j}}}  \mathcal{PDF}_{\rm{sig},\it{i}} \otimes G(\mu_{i}, \sigma_{i}) \nonumber \\
    && +  N_{\rm{bkg,fit}}  \frac{N_{\rm{bkg},\it{i}}}{\sum_j N_{\rm{bkg},\it{j}}}  \mathcal{PDF}_{\rm{bkg},\it{i}}),
\end{eqnarray}
where $i, j$ denote individual tag modes,
$N_{\rm{sig,fit}}$ and $N_{\rm{bkg,fit}}$ are the yields of signal and background events,
$\mathcal{PDF}_{\rm{sig},\it{i}} \otimes G(\mu_{i}, \sigma_{i})$ is
the probability density function derived from the signal MC distribution convolved with a Gaussian function~$G(\mu_{i}, \sigma_{i})$, where $\mu_{i}$ and $\sigma_{i}$ are obtained from the control samples.
The Gaussian function~$G(\mu_{i}, \sigma_{i})$ is introduced to
account for the signal resolution difference between data and MC simulation.
The $\mathcal{PDF}_{\rm{bkg},\it{i}}$ refers to the background shape
derived from the inclusive MC sample using \texttt{Meerkat} with
kernel density estimation~\cite{kernel}. $\mathcal{B}(D^0)_j$ represents the BF for each $D^0$ tag mode,
and $\epsilon_{\rm{sig},\it{i}}$ is the signal efficiency.
The distribution of the simultaneous fit is shown in the top left of Figure~\ref{fig:fit}, while the remaining three sub-figures illustrate the fit results for Modes~I, II, and III, individually.
The simultaneous fit yields $N_{\rm{sig}} = 0.1 \pm 13.5$, indicating that no significant signal is observed.

\vspace{-0.0cm}
\vspace{-0.0cm}

We then set the UL on $\mathcal{B}(\jpsi\to D^{0}\mu^{+}\mu^{-})$ at the $90\%$ C.L. after considering all systematic uncertainties.
The BF is calculated by
\begin{eqnarray}
\mathcal{B}(\jpsi \to D^{0}\mu^{+}\mu^{-}) = \frac{N_{\rm{sig,fit}} }{N_{J/\psi}\sum_{i}  \epsilon_{\rm{sig},\it{i}} \mathcal{B}_{\rm{inter},\it{i}}}.
\label{equation:Simultaneous BF}
\end{eqnarray}
where $N_{\jpsi} = (10087\pm44)\times10^{6}$ is the total number of $\jpsi$ events~\cite{bes3:totJpsiNumber},
$N_{\rm{sig,fit}}$ is the number of signal events obtained from the simultaneous fit,
$\epsilon_{\rm{sig},\it{i}}$ and $\mathcal{B}_{\rm{inter},\it{i}}$ are the signal efficiency and the BFs of the three $D^0$ decay channels from the PDG~\cite{pdg:2024}, respectively.

We scan the $\jpsi\to D^{0}\mu^{+}\mu^{-}$ signal yield 700 times by fixing the number of events from 0 to 70 with the step size of 0.1 to determine a signal-yield dependent likelihood function.
This function is described with a Gaussian distribution
\begin{eqnarray}
\mathcal{L}(\mathcal{B})_{\rm{fit}}\propto \rm{exp}[-\frac{(\mathcal{B}-\hat{\mathcal{B}})^{2}}{2\sigma^{2}_{\mathcal{B}}}],
\label{eq:B}
\end{eqnarray}
where $\hat{\mathcal{B}}$ is the mean value of BF and $\sigma_{\mathcal{B}}$ is the standard deviation of the Gaussian distribution.

To include the systematic uncertainties presented in the following section, we adopt the method in~Refs.~\cite{semarscan1, semarscan2} by smearing the likelihood distribution to
\begin{eqnarray}
\mathcal{L}(\mathcal{B})_{\rm{smear}}\propto \int_{0}^{1}\rm{exp}[-\frac{(\epsilon\mathcal{B}/\hat{\epsilon}-\hat{\mathcal{B}})^{2}}{2\sigma^{2}_{\mathcal{B}}}]\times\frac{1}{\sqrt{2\pi}\sigma_{\epsilon}}\rm{exp}[-\frac{(\epsilon-\hat{\epsilon})^{2}}{2\sigma^{2}_{\epsilon}}]d\epsilon,
\label{eq:B2}
\end{eqnarray}
where $\hat{\epsilon}$ is the average signal efficiency of the three tag modes, $\sigma_{\epsilon}=\Delta_{\rm{sys}}\cdot\hat\epsilon$ is the systematic uncertainty of efficiency.

\begin{figure*}[htbp]
\centering
	\setlength{\abovecaptionskip}{-1pt}
	\setlength{\belowcaptionskip}{10pt}
	\includegraphics[width=0.6\textwidth]{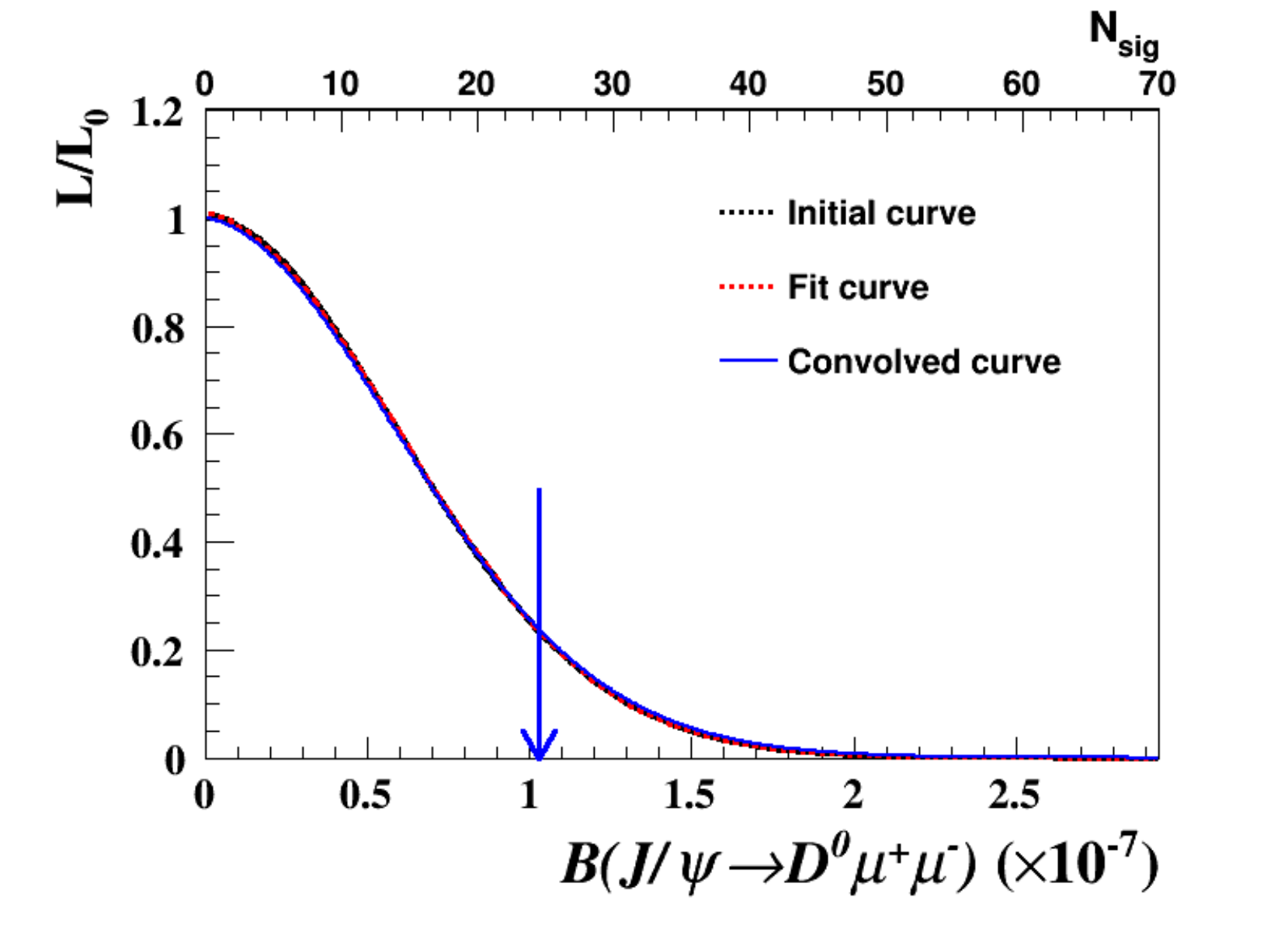}
	\caption{
 The distributions of normalized likelihoods versus signal yields $N_{\rm{sig}}$ or BF of $\jpsi\to D^{0}\mu^{+}\mu^{-}$.
 The black dashed curve represents the initial distribution,
 while the red dashed curve shows the fitted curve with a Gaussian function.
 The blue solid line illustrates the result convolved with a Gaussian function for systematic uncertainties.
 The blue arrow indicates the UL on its BF at the 90\% C.L.
 }
	\label{fig:UL}
\end{figure*}

The distributions of the likelihood curves are shown in Figure~\ref{fig:UL}.
The likelihoods on the $\mathcal{B}$ are integrated in the physical region ($\mathcal{B} > 0$), resulting in a UL to $\mathcal{B}(\jpsi\to D^{0}\mu^{+}\mu^{-})<1.1\times10^{-7}$ at the $90\%$ C.L.


\vspace{-0.0cm}
\vspace{-0.0cm}

\section{SYSTEMATIC UNCERTAINTY}
\label{sec:systematic}
\hspace{1.5em}
The systematic uncertainties in the BF measurement of $\jpsi\to D^{0}\mu^{+}\mu^{-} $ include
the signal MC model, the tracking and PID, the $D^0$ decay BFs, the total number of $\jpsi$ events, and the event selection criteria.
For the event selection criteria, data-MC discrepancies measured with
control samples are used to evaluate the uncertainties.
The systematic uncertainties from different sources are studied in the following items and summarized in Table~\ref{tab:syst_err}.

\begin{table*}[tpb]
\setlength{\abovecaptionskip}{0.0cm}
\setlength{\belowcaptionskip}{-1.6cm}
\caption{The systematic uncertainties for each $D^{0}$ tag mode.}
  \begin{center}
  \footnotesize
  \newcommand{\tabincell}[2]{\begin{tabular}{@{}#1@{}}#2\end{tabular}}
  \begin{threeparttable}
  \begin{tabular}{l|c c c c c c}
      \hline \hline
  		Source/Mode &I~($\%$)   &II~($\%$)   & III~($\%$) \\
		\hline
            Signal MC model                                       &\multicolumn{3}{c}{1.3}                 \\
            Number of $J/\psi$ events                               &\multicolumn{3}{c}{0.5}                   \\
            $E_{\mu}$                                         &\multicolumn{3}{c}{2.8}                        \\
		Tracking                                          & 4.0                      & 4.0                                     & 6.0\\
		PID                                       & 4.0                       & 4.0                                      & 6.0\\
            $\chi_{D^0}^2$                               & 5.8                       & 3.8                                       & 10.2\\
            $\pi^0$ reconstruction                            & -                           & 2.2                                      & -\\
            $M_{D^{0}}$                                       & 0.6                       & 0.4                                       & 1.3\\
		MC Statistics                           & 0.3                        & 0.5                                      & 0.5 \\
		Intermediate BFs                                  & 0.8                       & 3.5                                      & 1.8\\
            $|\vec{p}_{\rm{miss}}|$                                & 0.3                       & 0.8                                     & 0.6\\
  		$M^{\rm{Recoil}}_{\pi\mu\mu}$                          & 4.5                       & -                                          & -\\
            $M^{\rm{Recoil}}_{\pi\pi^0\mu\mu}$                     & -                           & 2.1                                      & -\\
            $M^{\rm{Recoil}}_{\pi\pi\pi\mu\mu}$                    & -                           & -                                          & 1.3\\
    	$M_{4\pi}$                                        & 0.7                       & -                                          & -\\
      	$M_{4\pi\pi^{0}}$                                 & -                           & 0.6                                       & -\\
            $M_{\pi^{\pm} \pi^0}$                                     & -                           & 0.4                                        & -\\
            $M_{\pi^{\pm} \pi^+\pi^-}$                                     & -                           & -                                        & 0.1\\
            $M_{\ks}$                                         & 0.3                       & 0.3                                      & 0.3\\
            \hline
             Total                                & 9.9                       & 9.0                                      & 13.9\\
            \hline
            Combinatorial total             &\multicolumn{3}{c}{9.7}   \\
		\hline\hline
  \end{tabular}
  \label{tab:syst_err}
  \end{threeparttable}
  \end{center}
\end{table*}

\begin{itemize}
    \item \emph{Signal MC model.}
    To estimate the systematic uncertainty of the signal model, we analyze the effective mass of the virtual vector particle using the signal MC model for different parameters~\cite{generator:D0ll}.
    The change of the signal efficiency, 1.3\%, is taken as the systematic uncertainty.

     \item \emph{Number of $\jpsi$ events.} The total number of $J/\psi$ events in data has been determined to be $N_{\jpsi}=(10087\pm44)\times10^{6}$ in Ref.~\cite{bes3:totJpsiNumber}, and its relative uncertainty of $0.5\%$ is taken as a systematic uncertainty.

     \item  \emph{$E_{\mu}$ selection.}
    The control sample of $\jpsi\to\mu^+\mu^-\gamma_{\rm{FSR}}$
    is utilized to estimate the systematic uncertainty of the $E_{\mu}$ requirement, which is assigned as 1.4\% per muon.

    \item \emph{Tracking and PID efficiencies.}
   For the tracking~(PID) efficiency for $K$ and $\pi$,
    the efficiency for detecting charged track is determined by analyzing doubly tagged $D^{0}\bar{D}^{0}$ events from $\psi(3770)$~\cite{wei:2016pidtrk}.
    The uncertainties of tracking~(PID) are assigned as 1.0\%~(1.0\%) per track.
    The tracking~(PID) uncertainty for $\mu$ is investigated by using a control sample of $\ee\to\gamma\mu^+\mu^-$~\cite{bes3:2017mutrk}, and is assigned to be 1.0\%~(1.0\%) per track.
    The systematic uncertainty of $\pi^{0}$ reconstruction is assigned as 2.2\%~\cite{BESIII:pi0}.
    The total systematic uncertainty of tracking~(PID) are 4.0\%, 4.0\%, and 6.0\%~(4.0\%, 4.0\%, and 6.0\%), respectively.

     \item  \emph{$M_{D^{0}}$ and $\chi_{D^0}^2$ selections.}
     To estimate the systematic uncertainties due to $M_{D^{0}}$ and
     $\chi_{D^0}^2$ selections, the control samples of $\psi(3770)\to D^{0}\bar{D}^0$ are chosen, with $D^{0} \to K^{-}\pi^{+}$, $D^{0} \to K^{-}\pi^{+}\pi^{0}$, and $D^{0} \to K^{-}\pi^{+}\pi^{+}\pi^{-}$ for Modes~I, II, and III, respectively.
    The systematic uncertainties for the $M_{D^{0}}$ ($\chi_{D^0}^2$) requirements are estimated to be 0.6\%, 0.4\%, and 1.3\% (5.8\%, 3.8\%, and 10.2\%), respectively.

     \item \emph{Limited signal MC sample.}
     The signal efficiencies are estimated from signal MC events, and the uncertainties of MC statistics are 0.3\%, 0.5\%, and 0.5\% for Modes~I, II, and III, respectively.

     \item \emph{The $D^0$ decay BFs.} The BFs of $D^{0} \to K^{-}\pi^{+}$, $D^{0} \to K^{-}\pi^{+}\pi^{0}$, and $D^{0} \to K^{-}\pi^{+}\pi^{+}\pi^{-}$ are taken from the PDG~\cite{pdg:2024}.
     The uncertainties are 0.8\%, 3.5\%, and 1.8\% for Modes~I, II, and III, respectively.

     \item The event selection criteria of \emph{$|\vec{p}_{\rm{miss}}|$, $M^{\rm{Recoil}}_{\pi\mu\mu}$, $M^{\rm{Recoil}}_{\pi\pi^0\mu\mu}$, $M^{\rm{Recoil}}_{\pi\pi\pi\mu\mu}$, $M_{4\pi}$, $M_{4\pi\pi^{0}}$, $M_{\pi^{\pm}\pi^0}$, and $M_{\pi^{\pm}\pi^+\pi^-}$.}
     To estimate the systematic uncertainties due to
     $|\vec{p}_{\rm{miss}}|$, $M^{\rm{Recoil}}_{\pi\mu\mu}$,
     $M^{\rm{Recoil}}_{\pi\pi^0\mu\mu}$,
     $M^{\rm{Recoil}}_{\pi\pi\pi\mu\mu}$, $M_{4\pi}$,
     $M_{4\pi\pi^{0}}$, $M_{\pi^{\pm}\pi^0}$, and
     $M_{\pi^{\pm}\pi^+\pi^-}$ selections, the control samples of
     $\jpsi \to K^{+}\pi^{+}\pi^{-}\pi^{-}$, $\jpsi \to
     K^{+}\pi^{+}\pi^{-}\pi^{-}\pi^{0}$, and~$\jpsi \to
     K^{+}\pi^{+}\pi^{+}\pi^{-}\pi^{-}\pi^{-}$ are chosen for Modes~I, II, and III, respectively.
    For Mode~I, the systematic uncertainties of the $|\vec{p}_{\rm{miss}}|$, $M^{\rm{Recoil}}_{\pi\mu\mu}$~and~$M_{4\pi}$ requirements are estimated to be 0.3\%, 4.5\%, and~0.7\%, respectively.
     For Mode~II, the systematic uncertainties of the $|\vec{p}_{\rm{miss}}|$, $M^{\rm{Recoil}}_{\pi\pi^0\mu\mu}$, $M_{4\pi\pi^{0}}$~and~$M_{\pi^{\pm}\pi^0}$ requirements are estimated to be 0.8\%, 2.1\%, 0.6\%, and~0.4\%, respectively.
     For Mode~III,  the systematic uncertainties of the $|\vec{p}_{\rm{miss}}|$, $M^{\rm{Recoil}}_{\pi\pi\pi\mu\mu}$,~and~$M_{\pi^{\pm}\pi^+\pi^-}$ requirements are estimated to be 0.6\%, 1.3\%, and~0.1\%, respectively.

     \item  \emph{$M_{\ks}$ veto.}
    The systematic uncertainty of the $M_{\ks}$ veto is estimated using a control sample of $\jpsi \to K^+ K^- \pi^+ \pi^-$.
    Due to strangeness conservation, this sample cannot include events from the decay $\ks \to \pi^+ \pi^-$.
    The systematic uncertainty is estimated to be~0.3\% for all $D^0$ tag modes.

       \item \emph{Background shape.}
    The systematic uncertainty of the background shape is evaluated by varying
    the background shape derived from the inclusive MC sample using \texttt{RooKeysPdf} method, third-order and fourth-order polynomial function, and the most conservative result is adopted as the final UL.

\end{itemize}

The square root of the quadratic sum of all the uncertainties is taken as the systematic uncertainty for each $D^0$ tag mode.
For Mode~I, II, and III, the total systematic uncertainties are 9.9\%, 9.0\%, and 13.9\%, respectively.

The system uncertainties in Table~\ref{tab:syst_err} are all multiplicative uncertainties, while the uncertainty for the background shape is additive uncertainty.
The combinatorial total multiplicative uncertainty for $\jpsi \to D^{0} \mu^{+}\mu^{-}$ is calculated by the error propagation:
\begin{eqnarray}
\Delta_{\mathcal{B},\rm{sys}} = \frac{1}{\mathcal{B}} \sqrt{\sum_{m, n} \rho_{mn} \frac{\partial \mathcal{B}}{\partial x_m} \frac{\partial \mathcal{B}}{\partial x_n} (\Delta_{x_m} x_m) (\Delta_{x_n} x_n}),
\label{equation:error propagation}
\end{eqnarray}
where $\mathcal{B}$ is the BF of $J/\psi \to D^{0} \mu^{+} \mu^{-}$,
$\Delta_{\mathcal{B},\rm{sys}}$ is the systematic uncertainty of the BF,
$x_m, x_n$ are defined as all the variables for the three tags,
$\Delta_{x_m}$, $\Delta_{x_n}$ are the systematic uncertainties of these variables,
and $\rho_{mn}$ represents the correlation coefficient between the variables $x_{m}$ and $x_{n}$.
If $m = n$, or if the two variables are correlated, then $ \rho_{mn} = 1 $.
In other cases, $ \rho_{mn} = 0 $.
In this paper, the uncertainty of signal MC models, number of $\jpsi$ events, $E_{\mu}$, tracking, PID, and $\chi_{D^0}^2$ between different modes are assigned as correlated.

The combinatorial total multiplicative systematic uncertainty is 9.7\%,
which has been used to calculate the UL in Section~\ref{sec:result}.

\section{SUMMARY}
\label{sec:summary}
\hspace{1.5em}
The FCNC decay $\jpsi \to D^{0}\mu^{+}\mu^{-} $ is searched for the first time based on $(10087\pm44)\times10^{6}$ $\jpsi$ events collected with the BESIII detector.
No significant signal is observed.
The UL on the BF is set to be $\mathcal{B}(\jpsi\to D^{0}\mu^{+}\mu^{-})<1.1\times10^{-7}$ at the $90\%$~C.L.
Due to the challenging selection of muons, this research does not achieve a better UL on the BF of $J/\psi\to D^0e^+e^-$~\cite{jpsi2D0ee}. 
However, this is the first search for a charmonium FCNC process involving muons in the final state,
and the result is compatible with the SM prediction on its BF of $10^{-13}$~\cite{FCNC2}.

\acknowledgments
\hspace{1.5em}
The BESIII Collaboration thanks the staff of BEPCII and the IHEP computing center for their strong support. This work is supported in part by National Key R\&D Program of China under Contracts Nos. 2023YFA1606000, 2020YFA0406400, 2020YFA0406300; National Natural Science Foundation of China (NSFC) under Contracts Nos. 11635010, 11735014, 11935015, 11935016, 11935018, 12025502, 12035009, 12035013, 12061131003, 12192260, 12192261, 12192262, 12192263, 12192264, 12192265, 12221005, 12225509, 12235017, 12361141819; the Chinese Academy of Sciences (CAS) Large-Scale Scientific Facility Program; the CAS Center for Excellence in Particle Physics (CCEPP); Joint Large-Scale Scientific Facility Funds of the NSFC and CAS under Contract No. U1832207; CAS under Contract No. YSBR-101; 100 Talents Program of CAS; The Institute of Nuclear and Particle Physics (INPAC) and Shanghai Key Laboratory for Particle Physics and Cosmology; Agencia Nacional de Investigación y Desarrollo de Chile (ANID), Chile under Contract No. ANID PIA/APOYO AFB230003; German Research Foundation DFG under Contract No. FOR5327; Istituto Nazionale di Fisica Nucleare, Italy; Knut and Alice Wallenberg Foundation under Contracts Nos. 2021.0174, 2021.0299; Ministry of Development of Turkey under Contract No. DPT2006K-120470; National Research Foundation of Korea under Contract No. NRF-2022R1A2C1092335; National Science and Technology fund of Mongolia; National Science Research and Innovation Fund (NSRF) via the Program Management Unit for Human Resources \& Institutional Development, Research and Innovation of Thailand under Contract No. B50G670107; Polish National Science Centre under Contract No. 2019/35/O/ST2/02907; Swedish Research Council under Contract No. 2019.04595; The Swedish Foundation for International Cooperation in Research and Higher Education under Contract No. CH2018-7756; U. S. Department of Energy under Contract No. DE-FG02-05ER41374.


\bibliographystyle{JHEP}
\bibliography{ref}

\newpage
M.~Ablikim$^{1}$, M.~N.~Achasov$^{4,c}$, P.~Adlarson$^{76}$, X.~C.~Ai$^{81}$, R.~Aliberti$^{35}$, A.~Amoroso$^{75A,75C}$, Q.~An$^{72,58,a}$, Y.~Bai$^{57}$, O.~Bakina$^{36}$, Y.~Ban$^{46,h}$
% ...
\\
\vspace{0.2cm}
(BESIII Collaboration)\\
\vspace{0.2cm} {\it
$^{1}$ Institute of High Energy Physics, Beijing 100049, People's Republic of China\\
% ...
$^{4}$ Budker Institute of Nuclear Physics SB RAS (BINP), Novosibirsk 630090, Russia\\
% ...
$^{35}$ Johannes Gutenberg University of Mainz, Johann-Joachim-Becher-Weg 45, D-55099 Mainz, Germany\\
$^{36}$ Joint Institute for Nuclear Research, 141980 Dubna, Moscow region, Russia\\
% ...
$^{46}$ Peking University, Beijing 100871, People's Republic of China\\
% ...
$^{57}$ Southeast University, Nanjing 211100, People's Republic of China\\
$^{58}$ State Key Laboratory of Particle Detection and Electronics, Beijing 100049, Hefei 230026, People's Republic of China\\
% ...
$^{72}$ University of Science and Technology of China, Hefei 230026, People's Republic of China\\
% ...
$^{75}$ University of Turin and INFN, (A)University of Turin, I-10125, Turin, Italy; (B)University of Eastern Piedmont, I-15121, Alessandria, Italy; (C)INFN, I-10125, Turin, Italy\\
$^{76}$ Uppsala University, Box 516, SE-75120 Uppsala, Sweden\\
% ...
$^{81}$ Zhengzhou University, Zhengzhou 450001, People's Republic of China\\

\vspace{0.2cm}
$^{a}$ Deceased\\
% ...
$^{c}$ Also at the Novosibirsk State University, Novosibirsk, 630090, Russia\\
% ...
$^{h}$ Also at State Key Laboratory of Nuclear Physics and Technology, Peking University, Beijing 100871, People's Republic of China\\
% ...

}
 \end{document}